\documentclass[reprint,amsmath,amssymb,aps,]{revtex4-2}

\usepackage{graphicx}
\usepackage{dcolumn}
\usepackage{bm}
\usepackage[]{hyperref}
\usepackage{color} 
\usepackage{amsmath}

\begin{document}

\title{Controlling Frequency-Domain Hong--Ou--Mandel Interference via Electromagnetically Induced Transparency}

\author{Zi-Yu Liu$^{1,2}$, Jiun-Shiuan Shiu$^{1,2}$, Chin-Yao Cheng$^{1,2}$, and Yong-Fan Chen$^{1,2,3}$}

\email{yfchen@mail.ncku.edu.tw}

\affiliation{$^1$Department of Physics, National Cheng Kung University, Tainan 70101, Taiwan \\
$^2$Center for Quantum Frontiers of Research $\&$ Technology, Tainan 70101, Taiwan \\
$^3$Center for Quantum Technology, Hsinchu 30013, Taiwan}

\date{\today}


\begin{abstract}

Hong--Ou--Mandel (HOM) interference is a compelling quantum phenomenon that demonstrates the nonclassical nature of single photons. In this study, we investigate an electromagnetically induced transparency-based double-$\Lambda$ four-wave mixing system from the perspective of quantized light fields. The system can be used to realize efficient HOM interference in the frequency domain. By using the reduced density operator theory, we demonstrate that although the double-$\Lambda$ medium does not exhibit phase-dependent properties for the closed-loop case of two incident single photons, frequency-domain HOM two-photon interference occurs. For experimentally achievable optical depth conditions, our theory indicates that this double-$\Lambda$ scheme can perform high-fidelity Hadamard gate operations on frequency-encoded single-photon qubits, and thereby generate HOM two-photon NOON states with a fidelity greater than 0.99. Furthermore, we demonstrate that this scheme can be used to realize arbitrary single-qubit gates and two-qubit SWAP gates by simply controlling the laser detuning and phase, exhibiting its multifunctional properties and providing a new route to scalable optical quantum computing.

\end{abstract}


\maketitle


\newcommand{\FigOne}{
    \begin{figure}[t]
    \centering
    \includegraphics[width = 8.6 cm]{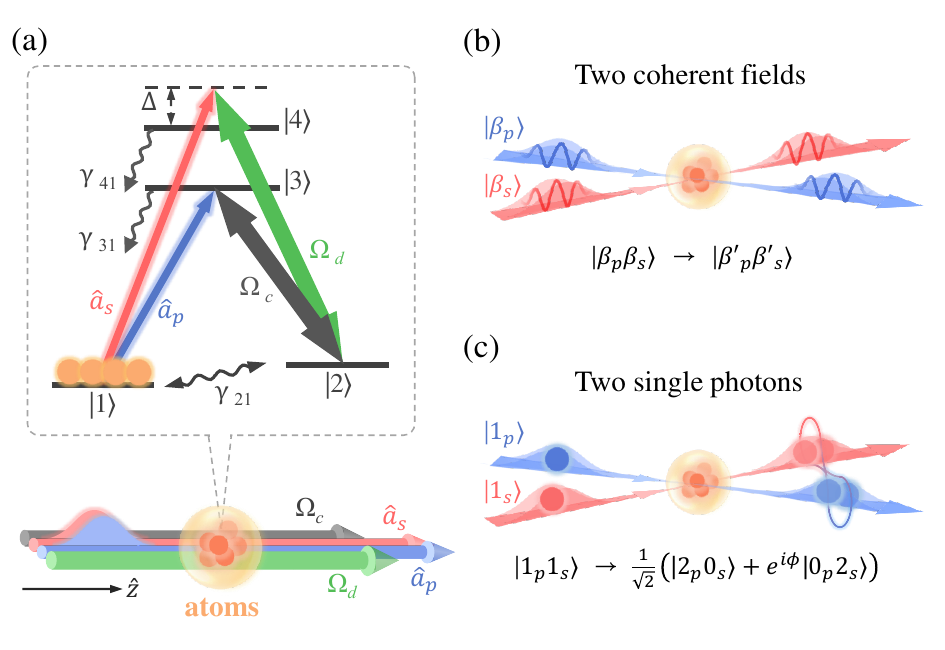}
    \caption{
Diagram of the four-level double-$\Lambda$ EIT system. (a) The corresponding transitions for the four applied light fields. All light fields are assumed to propagate in the same direction to eliminate the Doppler shift in the atomic medium. (b) Both the input probe and signal fields are in coherent states. In this case, the two coherent fields $|\beta_p\rangle$ and $|\beta_s\rangle$ interfere in the double-$\Lambda$ medium, and the output of the two coherent fields depends on the relative phase of the applied light fields. (c) Input probe and signal fields are two single photons (i.e., $|1_p1_s\rangle$). In this case, although the phase-dependent property of the double-$\Lambda$ medium is no longer present, HOM interference of the two-photon state occurs. The frequency-domain two-photon NOON state can be generated simply by optimizing the OD and $\Delta$.
}
    \label{fig:Energy level diagram}
    \end{figure}
}

\newcommand{\FigTwo}{
    \begin{figure}[t]
    \centering
    \includegraphics[width = 8.6 cm]{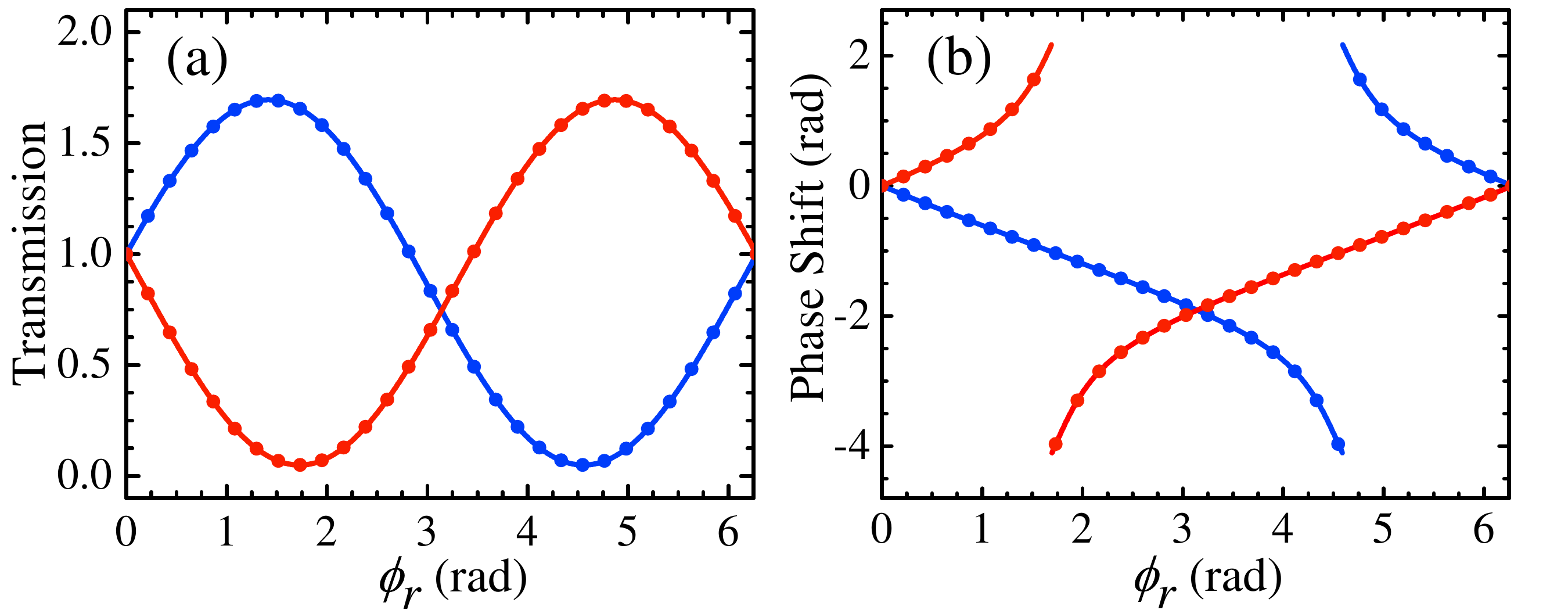}
    \caption{
The closed-loop case of two incident weak coherent fields. (a) Transmission and (b) phase shift of the probe (blue) and signal (red) coherent fields versus $\phi_r$ plotted using the quantum (solid) and semiclassical (dotted) models under conditions of $|\beta_p| = |\beta_s|$, $\Delta = 13\Gamma$, and $\textrm{OD} = 50$.
}
    \label{fig:Two coherent fields}
    \end{figure}
}

\newcommand{\FigThree}{
    \begin{figure}[t]
    \centering
    \includegraphics[width = 8.6 cm]{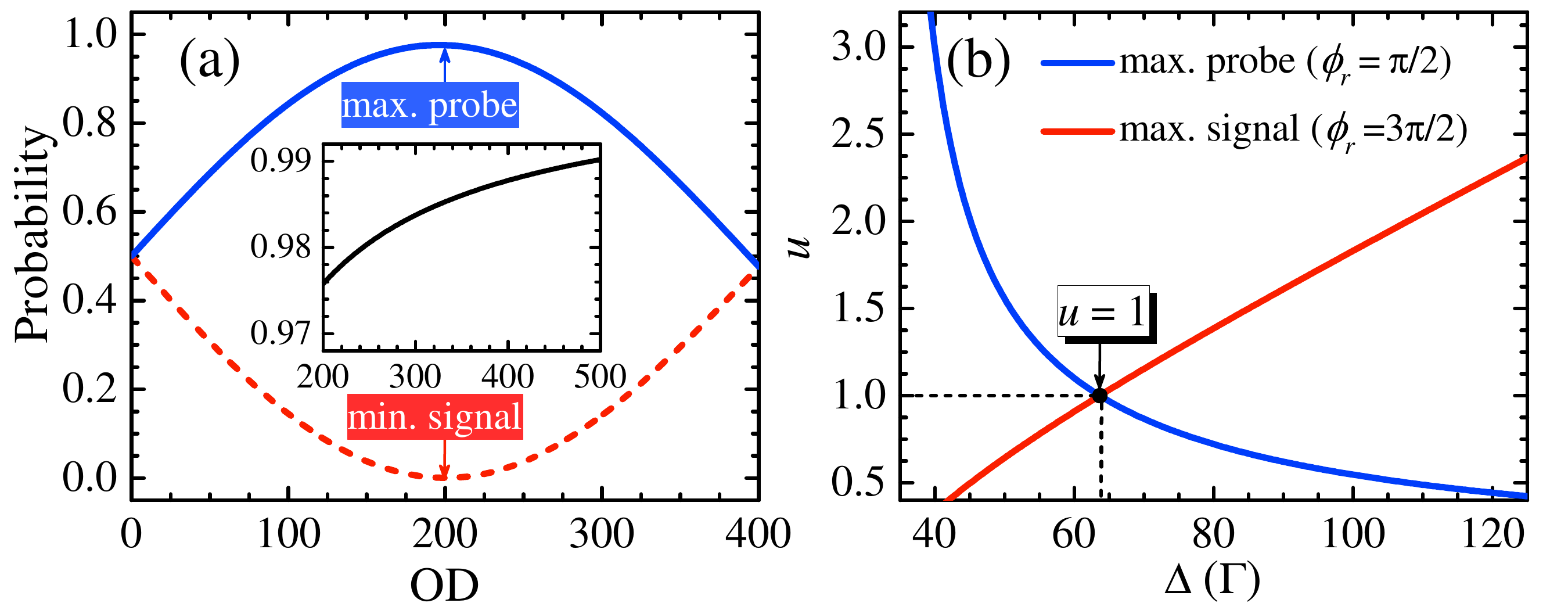}
    \caption{
Hadamard gates for single-photon two-color qubits. (a) Probability that the two-color qubit with $u=1$ and $\phi_u=0$ is in the probe frequency mode ($P_{1p0s}$, solid blue) and the signal frequency mode ($P_{0p1s}$, dashed red) versus OD under the conditions of $\phi_r = \pi/2$ and $\Delta = (200/\pi)\Gamma$. Inset: For $\Delta = (\textrm{OD}/\pi)\Gamma$, the maximum probability approaches 1 as OD increases. (b) Theoretical $\Delta$ curves for maximizing the probabilities of $|1_p0_s\rangle$ (solid blue) and $|0_p1_s\rangle$ (solid red) as a function of $u$ for $\textrm{OD} = 200$. The two curves intersect at $u=1$ where $\Delta = (200/\pi)\Gamma$, which is the condition in (a).
}
    \label{fig:Hadamard gates}
    \end{figure}
}

\newcommand{\FigFour}{
    \begin{figure}[t]
    \centering
    \includegraphics[width = 9.0 cm]{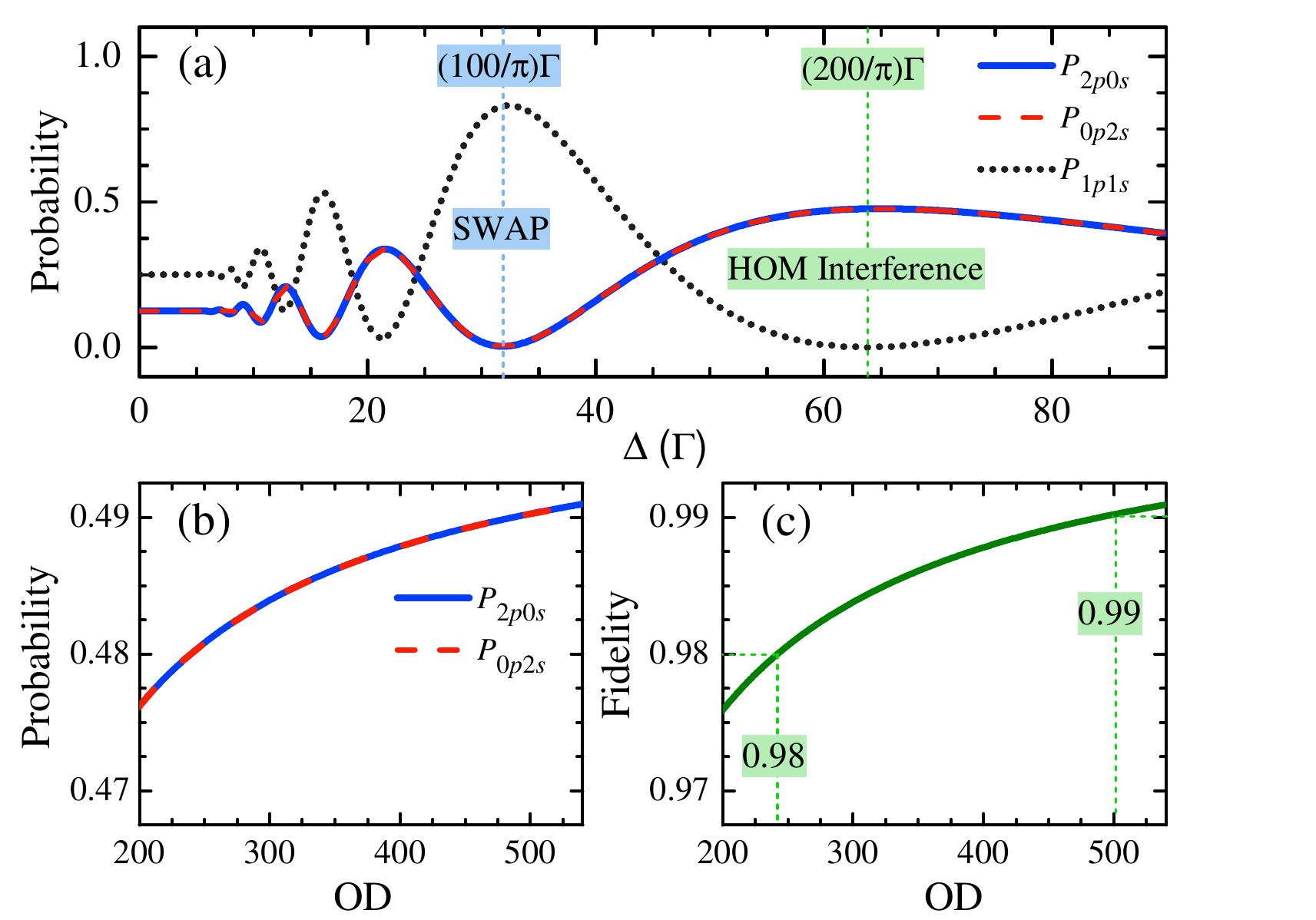}
    \caption{
HOM interference and color swap for two input single photons. (a) Theoretical curves for the probabilities $P_{2p0s}$ (solid blue), $P_{0p2s}$ (dashed red), and $P_{1p1s}$ (dotted black) versus $\Delta$ at $\textrm{OD} = 200$. HOM interference and two-photon swap occur at $\Delta = (200/\pi)\Gamma$ and $\Delta = (100/\pi)\Gamma$, respectively. (b) At $\Delta = (\textrm{OD}/\pi)\Gamma$, the maximum probabilities of $P_{2p0s}$ and $P_{0p2s}$ tend to approach 1 as OD increases. (c) The fidelity of the HOM two-photon NOON state versus OD.
}
    \label{fig:HOM interference}
    \end{figure}
}

\newcommand{\FigFive}{
	\begin{figure}[t]
		\centering
		\includegraphics[width = 8.6 cm]{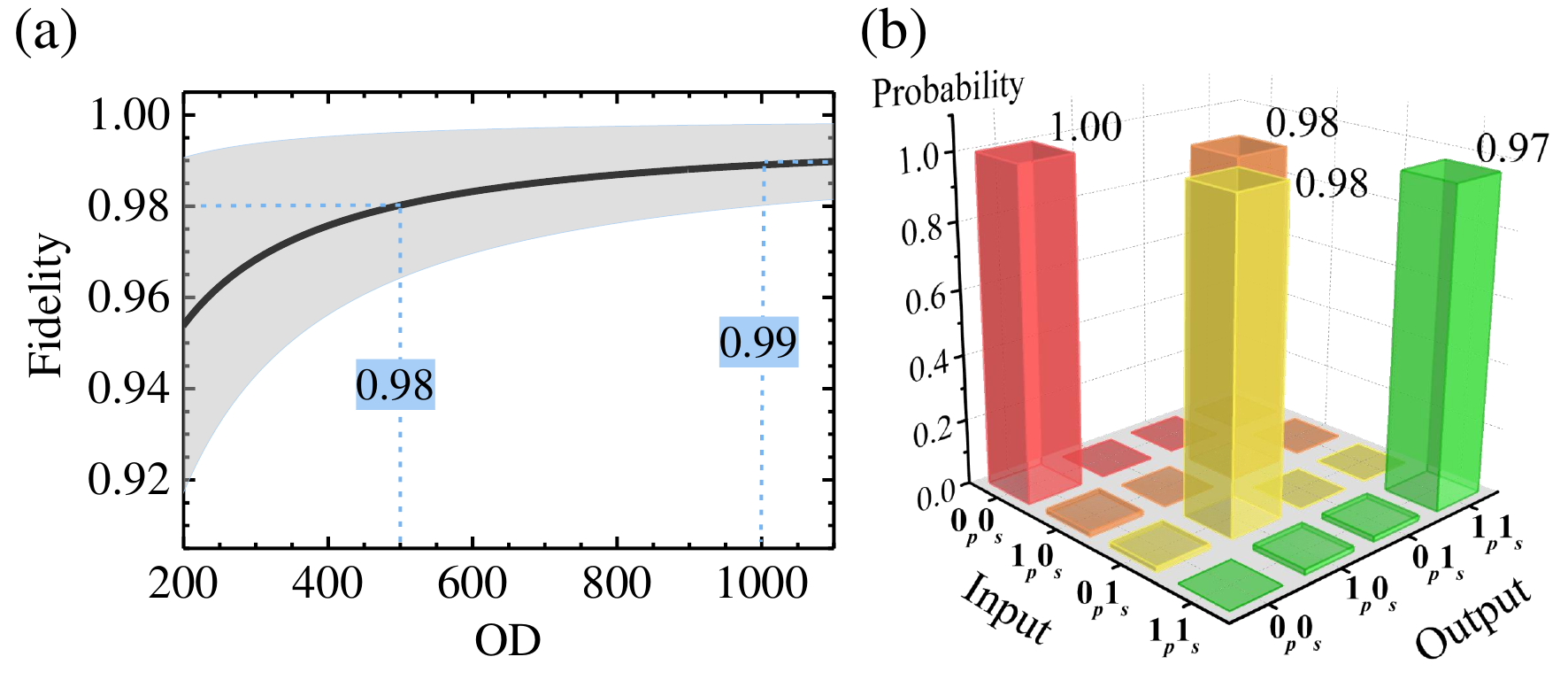}
		\caption{
Performance of the color SWAP gate. (a) Average (solid line) and standard deviation (gray area) of the SWAP gate fidelity plotted against the OD obtained from the density matrices of the output results according to the four computational input states. The vertical dash lines denote the OD of around $500$ and $1000$ for fidelities reaching  $0.98$ and $0.99$, respectively. (b) Truth table of the SWAP process according to the output probabilities using the four inputs when $\textrm{OD}=1000$ in (a). Under this condition, the average success rate per gate operation is around $0.98$.
}
    \label{fig:SWAP gates}
	\end{figure}
}


\section{Introduction} 

Hong--Ou--Mandel (HOM) interference~\cite{HOM1, HOM2, HOM3}, a well-known quantum effect without a classical counterpart, has played a key role in optical quantum information processing~\cite{OQP1, OQP2, OQP3, OQP4} and quantum metrology~\cite{OQM1, OQM2, OQM3, OQM4}. Instead of passive devices, such as spatial or polarizing beam splitters, recent studies have used frequency converters to transfer HOM interference from the spatial to the spectral domain; this method facilitates quantum frequency multiplexing and engineering~\cite{QFM1, QFM2, QFM3, QFM4}. Studies of frequency-domain HOM interference for single photons have relied on linear optics, which involve the use of electro-optic phase modulators~\cite{EPM1, EPM2, EPM3, Noise1}, or the nonlinear effects of solid materials~\cite{NSM1, NSM2, NSM3, Noise2}. Both schemes operate under far-detuned interaction conditions, thus typically requiring the use of high-voltage amplifiers or strong pump light, which leads to the generation of adjacent sidebands or parametric noise photons. Herein, we propose another promising scheme based on near-resonant nonlinear optics that enables frequency-domain HOM interference with high visibility. The proposed scheme leverages the closed-loop four-wave mixing (FWM) effect in a double-$\Lambda$ electromagnetically induced transparency (EIT) system~\cite{Fleischhauer}. Because of the EIT effect, this near-resonant FWM scheme can substantially suppress vacuum field noise at low light levels, thereby effectively preventing photon dissipation and avoiding the generation of noise photons~\cite{Cheng}. Such properties enable this double-$\Lambda$ scheme to generate frequency-domain HOM two-photon NOON states with extremely high fidelity.

To study frequency-domain HOM interference in a closed-loop double-$\Lambda$ medium, we constructed a quantum model based on the reduced density operator theory~\cite{ScullyQO}. Unlike semiclassical approaches, the proposed model quantizes two incident weak light fields. No discussion of such closed-loop FWM systems from the perspective of quantized light fields was found in a literature review. Previous semiclassical models have demonstrated that the respective transmittances of the two incident light fields through a closed-loop double-$\Lambda$ medium are affected by the relative phases of the applied light fields; this is known as the phase-dependent property~\cite{Korsunsky, Tsai, Artoni}. However, our quantum model revealed that the phase-dependent property of a closed-loop double-$\Lambda$ medium depends on the quantum states of the two incident weak light fields. If both incident weak light fields are in coherent states, the double-$\Lambda$ medium exhibits phase-dependent properties, which is consistent with the predictions of a semiclassical model and has been experimentally confirmed~\cite{ChenXPM, Stophel, Kang}. By contrast, if both incident weak light fields are single photons, this phase-dependent property is no longer present. Although the double-$\Lambda$ medium does not have phase-dependent properties in the case of two incident single photons, frequency-domain HOM interference of the two-photon state occurs. Our theory shows that such a double-$\Lambda$ scheme can perform high-fidelity Hadamard gate operations on frequency-encoded single-photon qubits. For experimentally achievable optical depth (OD) conditions, HOM two-photon NOON states with a fidelity of 0.99 can be generated. Furthermore, we demonstrate for the first time that this scheme can be used to implement arbitrary single-qubit gates and high-fidelity two-qubit SWAP gates, revealing its advantages in fabricating multifunctional photonic logic gates.

The paper is organized as follows. Section~\ref{sec:Theory} introduces the model of the double-$\Lambda$ EIT system, which incorporates quantized weak input fields. The quantum states of the output fields are derived using the reduced density operator theory. In Sec.~\ref{sec:Two cohenert fields}, we explore the input state consisting of two coherent fields and compare the results with the semiclassical model. Section~\ref{sec:Hadamard gates} investigates the utilization of a single-photon qubit and discusses the conditions for implementing the color Hadamard gate within the current system. The conditions for frequency-domain HOM interference with two input single photons are discussed in Sec.~\ref{sec:HOM interference}. Furthermore, Sec.~\ref{sec:SWAP gates} proposes and demonstrates the capability of the double-$\Lambda$ EIT system to generate two-qubit SWAP gates. Finally, the conclusions are presented in Sec.~\ref{sec:Conclusion}.


\section{Theoretical model} \label{sec:Theory}

We consider a medium comprising double-$\Lambda$ four-level atoms with two metastable ground states and two excited states [Fig.~\ref{fig:Energy level diagram}(a)]. The weak probe field $\hat{a}_p$ driving the $|1\rangle\leftrightarrow|3\rangle$ transition and the weak signal field $\hat{a}_s$ driving the $|1\rangle\leftrightarrow|4\rangle$ transition can both be described by quantum operators obeying the bosonic commutation relation. High-intensity coherent coupling and driving fields, described by the semiclassical Rabi frequency $\Omega_{c(d)}=|\Omega_{c(d)}|e^{i\phi_{c(d)}}$ with a well-defined phase $\phi_{c(d)}$, drive  $|2\rangle\leftrightarrow|3\rangle$ and $|2\rangle\leftrightarrow|4\rangle$ transitions, respectively, forming a closed-loop double-$\Lambda$ EIT system together with the probe and signal fields. The parameters $\gamma_{31}$ and $\gamma_{41}$ represent the total coherence decay rates of the excited states $|3\rangle$ and $|4\rangle$, respectively. $\gamma_{21}$ is the dephasing rate between the ground states $|1\rangle$ and $|2\rangle$. We consider the case in which the signal field and the driving field satisfy the two-photon resonance condition, but a one-photon detuning, denoted as $\Delta$, exists. By using the collective atomic operator approach and ignoring interactions between atoms, the interaction Hamiltonian $\hat{H}$ of the double-$\Lambda$ system can be expressed as follows:
\begin{align}
	\hat{H} & = -\frac{\hbar N}{2L}\int^L_0  \big[2g_p\hat{\sigma}_{31}(z,t)\hat{a}_p(z,t)+2g_s\hat{\sigma}_{41}(z,t)\hat{a}_s(z,t)
	\notag\\
	& + \Omega_c\hat{\sigma}_{32}(z,t)+\Omega_d \hat{\sigma}_{42}(z,t)+ \Delta\hat{\sigma}_{44}(z,t)+\textrm{h.c.} \big]dz,
	\label{eq:(1)}
\end{align}
where $N$ and $L$ are the total number of atoms and the atomic medium length, respectively. $g_{p}=\frac{d_{31} \varepsilon_{p}}{\hbar}$ denotes the coupling constant between the probe field and the single atom,  where $d_{31}$ is the dipole moment of the corresponding $|3\rangle\leftrightarrow|1\rangle$ transition and $\varepsilon_{p}=\sqrt{\frac{\hbar \omega_{p}}{2\epsilon_0 V}}$ is the electric field of the single probe photon. Similarly, there are $g_{s}=\frac{d_{41} \varepsilon_{s}}{\hbar}$ and $\varepsilon_{s}=\sqrt{\frac{\hbar \omega_{s}}{2\epsilon_0 V}}$ parameters for the signal field. $\hat{\sigma}_{jk}(z,t)$ represents a collective atomic operator in the slowly-varying amplitude approximation~\cite{FleischhauerQM} that obeys the following Heisenberg--Langevin equation (HLE) between states $|j\rangle$ and $|k\rangle$
\begin{eqnarray}
	\frac{\partial}{\partial t}\hat{\sigma}_{jk}=\frac{i}{\hbar}[\hat{H},\hat{\sigma}_{jk}]-\frac{\gamma_{jk}}{2}\hat{\sigma}_{jk}+\gamma^{sp}_{jk}+\hat{F}_{jk},
	\label{eq:(2)}
\end{eqnarray}
where $\gamma^{sp}_{jk}$ is the parameter related to the spontaneous decay process, and $\hat{F}_{jk}$ denotes the associated Langevin noise operator. In the context of the double-$\Lambda$ EIT system, the probe and signal fields can be considered as perturbation fields due to their significantly weaker strengths compared to the coupling and driving fields. Therefore, we can solve the first-order atomic operators by substituting the zero-order perturbation results into the relevant first-order HLEs as follows:
\begin{align}
	\frac{\partial}{\partial t} & \hat{\sigma}^{(1)}_{21}(z,t)  =
	\hat{F}_{21}(z,t)-\frac{1}{2}\gamma_{21}\hat{\sigma}^{(1)}_{21}(z,t) \notag\\
	& -i\left[ \frac{\Omega_c}{2}\hat{\sigma}^{(1)}_{31}(z,t)+\frac{\Omega_d}{2}\hat{\sigma}^{(1)}_{41}(z,t) \right], \label{eq:(3)} \\
	\frac{\partial}{\partial t} & \hat{\sigma}^{(1)}_{31}(z,t)  =
	\hat{F}_{31}(z,t)-\frac{1}{2}\gamma_{31}\hat{\sigma}^{(1)}_{31}(z,t) \notag\\
	& -i\left[ g_p\hat{a}^{\dagger}_p(z,t) +\frac{\Omega^{\ast}_c}{2}\hat{\sigma}^{(1)}_{21}(z,t) \right], \label{eq:(4)} \\
	\frac{\partial}{\partial t} & \hat{\sigma}^{(1)}_{41}(z,t)  =
	\hat{F}_{41}(z,t)-\frac{1}{2}\gamma_{41}\hat{\sigma}^{(1)}_{41}(z,t) \notag\\
	& -i\left[ g_s\hat{a}^{\dagger}_s(z,t) +\frac{\Omega^{\ast}_d}{2}\hat{\sigma}^{(1)}_{21}(z,t) +\Delta\hat{\sigma}^{(1)}_{41}(z,t) \right]. \label{eq:(5)}
\end{align}

\FigOne

To investigate the behavior of the probe and signal fields propagating in the double-$\Lambda$ medium, the solutions of $\hat{\sigma}^{(1)}_{31}(z,t)$ and $\hat{\sigma}^{(1)}_{41}(z,t)$ from the first-order HLEs are incorporated into the following Maxwell--Schr\"{o}dinger equations (MSEs):
\begin{align}
	\bigg(\frac{\partial}{\partial t}+c\frac{\partial}{\partial z}\bigg)\hat{a}^{\dagger}_{p}(z,t)=-ig_{p}N\hat{\sigma}^{(1)}_{31}(z,t), \label{eq:(6)} \\
	\bigg(\frac{\partial}{\partial t}+c\frac{\partial}{\partial z}\bigg)\hat{a}^{\dagger}_{s}(z,t)=-ig_{s}N\hat{\sigma}^{(1)}_{41}(z,t), \label{eq:(7)}
\end{align}
where $c$ is the speed of light. Subsequently, through the Fourier transform from the time domain to the frequency domain, the coupled equations of $\tilde{a}^{\dagger}_{p}(z, \omega)$ and $\tilde{a}^{\dagger}_{s}(z, \omega)$ can be derived as follows:
\begin{align}
	& \frac{\partial}{\partial z} \tilde{a}^{\dagger}_p
	+\Lambda_p \tilde{a}^{\dagger}_p +\kappa_p e^{i(-\phi_c+\phi_d)} \tilde{a}^{\dagger}_s =\sum_{jk}\zeta^p_{jk}\tilde{f}_{jk}, \label{eq:(8)} \\
	& \frac{\partial}{\partial z} \tilde{a}^{\dagger}_s
	+\Lambda_s \tilde{a}^{\dagger}_s +\kappa_s e^{-i(-\phi_c+\phi_d)} \tilde{a}^{\dagger}_p=\sum_{jk}\zeta^s_{jk}\tilde{f}_{jk}, \label{eq:(9)}
\end{align}
where $\Lambda_{p(s)}$ and $\kappa_{p(s)}$ are the EIT dispersion coefficient and coupling constant of the probe (signal) transition, respectively~\cite{Kolchin}. $\tilde{f}_{jk}\equiv\sqrt{\frac{N}{c}}\tilde{F}_{jk}$ is the renormalized Langevin noise operator, where $jk\in\lbrace 21, 31, 41\rbrace$ denotes the subscript of the atomic operator $\tilde{\sigma}_{jk}$. $\zeta^p_{jk}$ and $\zeta^s_{jk}$ are the coefficients of the Langevin noise operators of interest. For simplicity, we consider the steady-state case of $\omega$ = 0 and assume that $|\Omega_c| = |\Omega_d| \equiv |\Omega|$, $g_p = g_s \equiv g$, $\gamma_{31} = \gamma_{41} \equiv \Gamma$, and $\gamma_{21} = 0$. Under these conditions, the general solution of the aforementioned coupled equations can be obtained as follows:
\begin{align}
	\begin{bmatrix}
		\tilde{a}^{\dagger}_{p,L} \\
		\tilde{a}^{\dagger}_{s,L} \\
	\end{bmatrix}
	= e^{-ML}
	\begin{bmatrix}
		\tilde{a}^{\dagger}_{p,0} \\
		\tilde{a}^{\dagger}_{s,0} \\
	\end{bmatrix}
	+\sum_{jk}\int^L_0e^{-M(L-z)}
	\begin{bmatrix}
		\zeta^p_{jk} \\
		\zeta^s_{jk} \\
	\end{bmatrix}
	\tilde{f}_{jk}dz, \label{eq:(10)}
\end{align}
where $\tilde{a}^{\dagger}_{p(s),L}$ and $\tilde{a}^{\dagger}_{p(s),0}$ are abbreviations for $\tilde{a}^{\dagger}_{p(s)}(L,0)$ and $\tilde{a}^{\dagger}_{p(s)}(0,0)$, respectively. All the coefficients of the general solution can be found in Appendix \ref{sec:Appendix A}. The term $e^{-ML}$ characterizes the parametric evolution of the double-$\Lambda$ medium. Its matrix form is
\begin{align}
	e^{-ML}
	& = \textrm{exp}\left(-
	\begin{bmatrix}
		\Lambda_p & \kappa_pe^{i(-\phi_c+\phi_d)} \\
		\kappa_se^{-i(-\phi_c+\phi_d)} & \Lambda_s \\
	\end{bmatrix}
	L\right)
	\notag\\
	& 
	\equiv
	\begin{bmatrix}
		A & B \\
		C & D \\
	\end{bmatrix}
	. \label{eq:(11)}
\end{align}
The second term on the right-hand side of Eq. (\ref{eq:(10)}) describes the quantum fluctuation effect induced by the vacuum field reservoir. To enable correspondence with more commonly used parameters, we use the substitution $\frac{\alpha\Gamma}{4L}=\frac{g^2N}{c}$, where $\alpha$ denotes the OD of the atomic medium. Therefore, the matrix elements in Eq. (\ref{eq:(11)}) are obtained as
\begin{align}
	A & = \frac{1}{2} \left[1+e^{i\alpha\Gamma/2(\Delta-i\Gamma)}\right], \label{eq:(12)} \\
	B & = \frac{1}{2} \left[1-e^{i\alpha\Gamma/2(\Delta-i\Gamma)}\right] e^{-i(\phi_c-\phi_d)}, \label{eq:(13)} \\
	C & = \frac{1}{2} \left[1-e^{i\alpha\Gamma/2(\Delta-i\Gamma)}\right] e^{i(\phi_c-\phi_d)}, \label{eq:(14)} \\
	D & = \frac{1}{2} \left[1+e^{i\alpha\Gamma/2(\Delta-i\Gamma)}\right]. \label{eq:(15)}
\end{align}
In Eq. (\ref{eq:(10)}), $A$ and $C$ ($D$ and $B$) represent the mode-preserved and mode-converted coefficients, respectively, of the probe (signal) field propagating in the double-$\Lambda$ medium. These coefficients are similar to the transmission and reflection coefficients of light passing through a spatial beam splitter.

To investigate the response of the double-$\Lambda$ medium to different input photon quantum states, the density matrix of the output state, $\rho_f=\tilde{U}\rho_i\tilde{U}^{\dagger}$, is constructed in accordance with the density matrix of the input state $\rho_i$ and the evolution operator $\tilde{U}$ of the combined system in the frequency domain, where $\rho_i=\rho^{PS}(z = 0, \omega)\otimes\rho^R$ is in the product form of the incident probe and signal fields $\rho^{PS}$ and the reservoir $\rho^R$. According to the reduced density operator approach employed in our previous works~\cite{Cheng, Hsu}, under the steady-state condition, the matrix element of the output probe and signal fields in the Fock-state basis can be expressed as:
\begin{align}
	\rho^{PS}_{m_pm_sn_pn_s}(L,0)
	& = \textrm{Tr}\left\lbrace \tilde{\rho}_{m_pm_sn_pn_s}(L,0) \rho_i\right\rbrace,
	\label{eq:(16)}
\end{align}
where $\tilde{\rho}_{m_pm_sn_pn_s}(L,0)$ = $\tilde{U}^{\dagger}\left(|n_p n_s\rangle\langle m_p m_s|\otimes I_R\right)\tilde{U}$ is the operator for the matrix element of the output probe and signal fields. If Eq. (\ref{eq:(10)}) is applied and the vacuum density matrix with annihilation and creation operators are rewritten as $|0\rangle\langle0|=\sum^{\infty}_{l=0}\frac{(-1)^l}{l!}(\tilde{a}^{\dagger})^l(\tilde{a})^l$~\cite{LouisellQO}, the operator $\tilde{\rho}_{m_pm_sn_pn_s}$ can be derived as
\begin{align}
	& \tilde{\rho}_{m_pm_sn_pn_s}(L,0)
	\notag\\
	& = \sum^{\infty}_{l_p,l_s=0}\chi_{mnl}
	(A \tilde{a}^{\dagger}_{p,0}+ B \tilde{a}^{\dagger}_{s,0})^{l_p + n_p} (C \tilde{a}^{\dagger}_{p,0}+D \tilde{a}^{\dagger}_{s,0})^{l_s + n_s}
	\notag\\
	& \times (A^{\ast} \tilde{a}_{p,0}+B^{\ast} \tilde{a}_{s,0})^{l_p + m_p} (C^{\ast} \tilde{a}_{p,0}+D^{\ast} \tilde{a}_{s,0})^{l_s + m_s},
	\label{eq:(17)}
\end{align}
where $\chi_{mnl}=\frac{1}{\sqrt{m_p!m_s!n_p!n_s!}}\frac{(-1)^{l_p+l_s}}{l_p!l_s!}$. In the derivation of Eq. (\ref{eq:(17)}), the normal-ordered noise caused by the Langevin operator $\tilde{f}_{jk}$ is proportional to the atomic population of the excited state~\cite{Lauk}; hence, it can be ignored under the perturbation condition of the weak probe and signal fields (see Appendix \ref{sec:Appendix B} for more details).


\section{Two coherent fields} \label{sec:Two cohenert fields}

We first consider the case of two coherent fields passing through the double-$\Lambda$ medium. As indicated in Fig.~\ref{fig:Energy level diagram}(b), the two coherent states with probe and signal frequency modes, respectively, are the initial states $|\psi^{PS}_{i}\rangle = |\beta_p\rangle \otimes |\beta_s\rangle$. If Eqs. (\ref{eq:(16)}) and (\ref{eq:(17)}) are calculated, the matrix element of the the output probe and signal fields can be obtained as
\begin{align}
	& \rho^{PS}_{m_pm_sn_pn_s}(L,0) \notag \\
	& = \sum^{\infty}_{l_p,l_s=0}\chi_{mnl}
	(A\beta^{\ast}_p+B\beta^{\ast}_s)^{l_p + n_p}(C\beta^{\ast}_p+D\beta^{\ast}_s)^{l_s + n_s}
	\notag \\
	& \quad \times(A^{\ast}\beta_p+B^{\ast}\beta_s)^{l_p + m_p}(C^{\ast}\beta_p+D^{\ast}\beta_s)^{l_s + m_s}
	\notag\\
	& = e^{-|A^{\ast}\beta_p+B^{\ast}\beta_s|^2}\frac{(A^{\ast}\beta_p+B^{\ast}\beta_s)^{m_p}(A\beta^{\ast}_p+B\beta^{\ast}_s)^{n_p}}{\sqrt{m_p!n_p!}}
	\notag\\
	& \quad \times e^{-|C^{\ast}\beta_p+D^{\ast}\beta_s|^2}\frac{(C^{\ast}\beta_p+D^{\ast}\beta_s)^{m_s}(C\beta^{\ast}_p+D\beta^{\ast}_s)^{n_s}}{\sqrt{m_s!n_s!}}, \label{eq:(18)}
\end{align}
by which the output state of the probe and signal fields becomes
\begin{align}
	|\psi^{PS}_{f}\rangle & = |A^{\ast}\beta_p+B^{\ast}\beta_s \rangle \otimes |C^{\ast}\beta_p+D^{\ast}\beta_s \rangle \notag\\
	& \equiv |\beta'_p\rangle \otimes |\beta'_s\rangle, \label{eq:(19)}
\end{align}
where both the output probe $|\beta'_p\rangle$ and signal $|\beta'_s\rangle$ fields remain in coherent states, indicating that the double-$\Lambda$ medium only coherently redistributes the incident probe and signal amplitudes through the EIT-based FWM process without changing the photon statistics of these two coherent fields or entangling them~\cite{Chang}. This makes the case of two coherent fields unable to produce HOM two-photon interference. Furthermore, Eq. (\ref{eq:(19)}) reveals that the amplitude and phase changes of the output probe and signal fields depend on the relative phase of the applied light fields. The transmittances and phase shifts of the probe and signal fields through the double-$\Lambda$ medium are given by
\begin{align}
	T_p & = \left|\frac{\beta'_p}{\beta_p}\right|^2 = \left|A^{\ast} + u_\beta B'^{\ast} e^{-i\phi_r}\right|^2, \label{eq:(20)} \\
	\Delta\phi_p & = \textrm{arg}\left[\frac{\beta'_p}{\beta_p}\right] = \textrm{arg}\left[ A^{\ast} + u_\beta B'^{\ast} e^{-i\phi_r}\right], \label{eq:(21)} \\
	T_s & = \left|\frac{\beta'_s}{\beta_s}\right|^2 = \left|u^{-1}_\beta C'^{\ast} e^{i\phi_r} + D^{\ast}\right|^2, \label{eq:(22)} \\
	\Delta\phi_s & =  \textrm{arg}\left[\frac{\beta'_s}{\beta_s}\right] = \textrm{arg}\left[u^{-1}_\beta C'^{\ast} e^{i\phi_r} + D^{\ast}\right], \label{eq:(23)}
\end{align}
where $B'^{\ast} \equiv B^{\ast} e^{-i(\phi_c - \phi_d)}$, $C'^{\ast} \equiv C^{\ast} e^{i(\phi_c-\phi_d)}$, $u_\beta \equiv |\beta_s|/|\beta_p|$ is the ratio of the probe and signal amplitudes, and $\phi_r \equiv (\phi_p-\phi_s)-(\phi_c-\phi_d)$ is the relative phase of the applied light fields. Figure~\ref{fig:Two coherent fields} reveals that the theoretical curves of transmittance and phase shift, as a function of the relative phase $\phi_r$ for the probe and signal fields obtained from the current quantum model (solid lines), are identical to those of the semiclassical model (dotted lines) in Ref.~\cite{ChenXPM}. If the input light fields are all in coherent states, this closed-loop double-$\Lambda$ medium exhibits phase-dependent properties, which have been experimentally investigated in atomic ensembles~\cite{ChenXPM, Stophel, Kang}.

\FigTwo


\section{Hadamard gates} \label{sec:Hadamard gates}

We next consider a single-photon qubit case in which an incident single photon has entangled probe and signal frequency modes, namely
\begin{eqnarray}
	|\psi^{PS}_{i}\rangle = \frac{1}{\sqrt{1+u^2}}\left(|1_p0_s\rangle + u e^{-i\phi_u}  |0_p1_s\rangle \right), \label{eq:(24)}
\end{eqnarray}
where $u$ is a parameter indicating the amplitude ratio between the probe and signal frequency modes, and $\phi_u$ denotes the relative phase between the two modes. A single photon with a superposition state of two frequency modes is also called a frequency-encoded photonic qubit or two-color qubit~\cite{Lee}. Here, $\phi_u$ is similar to the role of the relative phase $(\phi_p-\phi_s)$ between the two coherent fields mentioned in the previous section. Therefore, in the case of a single-photon two-color qubit, the double-$\Lambda$ medium can also be expected to exhibit phase-dependent properties. To confirm this, the probability that the output single photon is in the probe and signal frequency modes, respectively, can be determined using Eqs. (\ref{eq:(16)}) and (\ref{eq:(17)}) as follows:
\begin{align}
	& P_{1p0s}=\rho^{PS}_{1010} = \frac{1}{1+u^2} \left |A^{\ast} + u B'^{\ast} e^{-i\phi_r} \right|^2, \label{eq:(25)} \\
	& P_{0p1s}=\rho^{PS}_{0101} = \frac{1}{1+u^2} \left |C'^{\ast} e^{i\phi_r} + u D^{\ast} \right|^2, \label{eq:(26)}
\end{align}
where $\phi_r = \phi_u-(\phi_c-\phi_d)$ is equivalent to the relative phase in the previous case of two coherent fields with $\phi_u = \phi_p-\phi_s$. As expected, these results verify the dependence of the output state of the single-photon qubit on the relative phase of the applied light fields in the double-$\Lambda$ medium. Figure~\ref{fig:Hadamard gates}(a) plots the theoretical curves of $P_{1p0s}$ and $P_{0p1s}$ as a function of OD under the conditions of $\phi_r =  \pi/2$ and $\Delta = \left(200/\pi\right)\Gamma$ for single-photon qubits with equal-amplitude and in-phase modes, namely $u = 1$ and $\phi_u = 0$. The theoretical curves reveal that $P_{1p0s}$ increases as OD increases and reaches a maximum value close to 1 at $\textrm{OD}$ = 200, whereas $P_{0p1s}$ decreases to almost zero. In other words, under these conditions, the two-color single photon entering the double-$\Lambda$ medium is converted into a single photon carrying only one color, which is equivalent to the Hadamard gate, and its operation fidelity can reach 0.99 at $\textrm{OD} = 200$.

\FigThree

The physical mechanism behind this frequency-domain Hadamard gate can be described as follows. Because of the large detuning condition of $\Delta = \left(\textrm{OD}/\pi\right)\Gamma$, the amplitudes of the mode-preserved coefficient and mode-converted coefficient of the double-$\Lambda$ medium are almost equal; that is, $|A| \approx |C| \approx 0.5$ and $|D| \approx |B| \approx 0.5$ [Eqs. (\ref{eq:(12)})--(\ref{eq:(15)})]. In this case, when $\phi_r = \pi/2$, the probe mode-preserved coefficient $A^{\ast}$ and the signal mode-converted coefficient $B'^{\ast} e^{-i\pi/2}$ in Eq. (\ref{eq:(25)}) are almost in phase, resulting in constructive interference, which causes the output photon to be converted into the probe frequency mode (i.e., $P_{1_p0_s} \approx$ 1). However, if $\phi_r$ is $3\pi/2$, the two aforementioned coefficients have out-of-phase destructive interference, causing the theoretical curves in Fig.~\ref{fig:Hadamard gates}(a) to become interchanged, and the output photon is converted to the signal frequency mode, resulting in $P_{0_p1_s} \approx$ 1. In addition, it is worth noting that under the condition of a sufficiently large OD and the specific detuning mentioned above, we can derive the unitary matrix of the gate operation using Eqs. (\ref{eq:(12)})--(\ref{eq:(15)}). By introducing a $\pi$ phase plate at the output of the signal field, the derived unitary matrix can be transformed into the unitary matrix representation of the Hadamard gate. Such results also indicate that the current scheme is applicable to any input state of a single-photon qubit.

For the input single-photon qubits with arbitrary $u$ values, the output of $|1_p0_s\rangle$ can be achieved by using a specific detuning value $\Delta$ that satisfies the $|A^{\ast}| = |uB'^{\ast}|$ condition. In this case, constructive interference in the probe frequency mode occurs if $\phi_r = \pi/2$. By contrast, the output of $|0_p1_s\rangle$ can be achieved by simply setting the relative phase $\phi_r$ to $3\pi/2$ and adjusting $\Delta$ to satisfy the condition $|C'^{\ast}| = |uD ^{\ast}|$. This suggests that the double-$\Lambda$ scheme can be used as a single-qubit gate with x-rotation and y-rotation. Therefore, arbitrary single-qubit gates can be realized by concatenating the x- and y-rotation gates based on this scheme~\cite{Nielsen}. Figure~\ref{fig:Hadamard gates}(b) presents plots of two theoretical curves of $\Delta$ as a function of $u$ required to achieve the maximum probabilities of $|1_p0_s\rangle$ and $|0_p1_s\rangle$ for $\textrm{OD} = 200$. The two curves intersect at $u = 1$ and $\Delta = (200/\pi)\Gamma$, corresponding to the condition $\textrm{OD} = 200$ of Fig.~\ref{fig:Hadamard gates}(a). For one-color input single photons ($u = 0$), the double-$\Lambda$ medium does not have phase-dependent properties because no interference occurs [Eqs. (\ref{eq:(25)}) and (\ref{eq:(26)})]. This system is also known as an open-loop FWM system and can be used as an efficient frequency converter~\cite{OFC1, OFC2, OFC3, OFC4, OFC5}.


\section{HOM interference} \label{sec:HOM interference}

We now consider the case in which the two input single photons carry the probe and signal frequency modes, respectively. This is a two-photon state with no correlation between the modes, as depicted by $|1_p\rangle \otimes |1_s\rangle$ in Fig.~\ref{fig:Energy level diagram}(c). As previously stated, under the ideal conditions of an extremely large OD value, the double-$\Lambda$ medium causes little photon dissipation. Therefore, the output photons only have the following three possible quantum states with corresponding probabilities: (1) state $|2_p0_s\rangle$ , both photons exit under the probe frequency mode with $P_{2p0s}$; (2) state $|0_p2_s\rangle$, both photons exit under the signal frequency mode with $P_{0p2s}$; (3) state $|1_p1_s\rangle$, photons exit with disparate colors with $P_{1p1s}$. These three probabilities are obtained as
\begin{align}
	& P_{2p0s} = \rho^{PS}_{2020} = 2\left|A^{\ast} B^{\ast} \right|^2, \label{eq:(27)} \\
	& P_{0p2s} = \rho^{PS}_{0202} = 2\left|C^{\ast} D^{\ast} \right|^2, \label{eq:(28)} \\
	& P_{1p1s} = \rho^{PS}_{1111} = \left|A^{\ast} D^{\ast} + B^{\ast} C^{\ast}\right|^2. \label{eq:(29)}
\end{align}
Unlike the previous case of the input two-color qubit, the probabilities of the output states in this case are not affected by the relative phase of the applied light fields; they are phase-independent. This phenomenon is ascribed to the quantum uncertainty of the relative phase between the two input single photons, which prevents them from interfering with each other in the double-$\Lambda$ medium. However, quantum interference of the two-photon state occurs [Eqs. (\ref{eq:(27)})--(\ref{eq:(29)})]. 

\FigFour

Figure~\ref{fig:HOM interference}(a) presents plots of the probabilities $P_{2p0s}$,  $P_{0p2s}$, and $P_{1p1s}$ versus $\Delta$ at $\textrm{OD}=200$. These theoretical curves reveal that if $\Delta = (\textrm{200}/\pi)\Gamma$, both $P_{2p0s}$ and $P_{0p2s}$ are close to 0.5, whereas $P_{1p1s}$ approaches 0. That is, the input two-photon state $|1_p1_s\rangle$ is almost completely suppressed by the double-$\Lambda$ medium, and the two single photons tend to have the same color, exhibiting HOM interference in the frequency domain. This phenomenon of two-photon state interference is caused by the operation of the Hadamard gate. If $\Delta = (\textrm{OD}/\pi)\Gamma$, the magnitudes of the coefficients for mode preservation and mode conversion are almost equal (approximately 0.5), which is the Hadamard gate operation and causes the transition amplitudes $A^{\ast}D^{\ast}$ (describing the two photons preserved in their respective modes) and $B^{\ast}C^{\ast}$ (indicating that the modes of the two photons mutually swap) to have the same magnitude but be out of phase (i.e., $A^{\ast}D^{\ast} \approx -B^{\ast}C^{\ast}$). Therefore, destructive interference between the two transition amplitudes in Eq. (\ref{eq:(29)}) occurs, resulting in $P_{1p1s}\approx 0$.

When the laser is detuned as $(\textrm{100}/\pi) \Gamma$ in Fig.~\ref{fig:HOM interference}(a), the two single photons exit with disparate colors; the state is $|1_p1_s\rangle$. In contrast with the two-photon HOM interference produced in the Hadamard gate operation, the mode-converted coefficients $B^{\ast}$ and $C^{\ast}$ are both approximately 1 in this case, whereas the mode-preserved coefficients $A^{\ast}$ and $D^{\ast}$ approach 0, resulting in $P_{1p1s}\approx 1$ and $P_{2p0s} \approx P_{0p2s} \approx 0$. Thus, the two single photons mutually swap colors in a near-perfect frequency conversion process and end up in the same state $|1_p1_s\rangle$, which also shows that whether the output photons generate HOM interference or swap colors can be controlled in this double-$\Lambda$ scheme by simply adjusting the laser detuning conditions.

Figure~\ref{fig:HOM interference}(b) plots the theoretical curves of $P_{2p0s}$ and $P_{0p2s}$ versus OD under the HOM interference condition of $\Delta$ = $\left(\textrm{OD}/\pi\right)\Gamma$. The theoretical results indicate that increasing OD can suppress the loss of output photons due to spontaneous emission. Therefore, if the OD value is large, the loss of output photons can be neglected, and its quantum state is obtained as
\begin{align}
	|\psi^{PS}_f\rangle =\frac{1}{\sqrt{2}}\left[|2_p0_s\rangle +e^{-2i(\phi_c-\phi_d)}|0_p2_s\rangle\right], \label{eq:(30)}
\end{align}
which is a two-photon NOON state in the frequency domain that has practical applications in photonic quantum metrology for its ability to conduct precision phase measurements if used in optical interferometers~\cite{Slussarenko}. Although the amplitude of the two-photon NOON state is independent of the phases of the applied light fields, the relative phase between modes $|2_p0_s\rangle$ and $|0_p2_s\rangle$ can be controlled by varying $\phi_c$ and $\phi_d$. Furthermore, for actual cases of finite OD, the two input single photons in the double-$\Lambda$ medium will be slightly dissipated, which not only causes the output photons to become a mixed state but also reduces the fidelity of the two-photon NOON state, as displayed in Fig.~\ref{fig:HOM interference}(c). The theoretical curve reveals that the fidelity approaches 1 as the OD increases owing to the proportional suppression of photon loss due to spontaneous emission in the double-$\Lambda$ EIT medium. If $\textrm{OD} = 500$, the fidelity can reach 0.99, which is a condition that can currently be achieved experimentally~\cite{OD1, OD2, OD3}. Hence, HOM interference based on the double-$\Lambda$ scheme is applicable for generating high-purity two-photon NOON states in the frequency domain.

\FigFive


\section{SWAP gates} \label{sec:SWAP gates}

In the previous section, we have shown that when the condition $\Delta =\left(\textrm{OD}/2\pi\right)\Gamma$ is satisfied, the colors of two input single photons are exchanged in the double-$\Lambda$ medium [see Fig.~\ref{fig:HOM interference}(a)]. Next, we demonstrate that this double-$\Lambda$ scheme can be used as a high-fidelity two-qubit SWAP gate under this condition, which is essential for scalable quantum networks~\cite{SQN1, SQN2}. Figure~\ref{fig:SWAP gates}(a) presents the average (curve) and standard deviation (gray area) of the SWAP gate fidelity according to the four computational input states (i.e., $|0_p0_s\rangle$, $|1_p0_s\rangle$, $|0_p1_s\rangle$, and $|1_p1_s\rangle$) versus \textrm{OD}. The theoretical curve shows that the average fidelity approaches 1 as the OD increases. By contrast, the standard deviation of fidelity decreases with increasing OD. When $\textrm{OD}=500$, the fidelity of the SWAP gate can reach $0.98$. For the region of \textrm{OD} exceeding $1000$, the SWAP gate achieves a high fidelity of $F=0.99$ with a low standard deviation of $0.01$. The truth table of the SWAP process when $\textrm{OD}=1000$ is shown in Fig.~\ref{fig:SWAP gates}(b). Under this condition, the average success probability per gate operation reaches $0.98$, corresponding to an error rate of $0.02$ for a single SWAP operation.


\section{Conclusion} \label{sec:Conclusion}

We have investigated a novel double-$\Lambda$ FWM system based on EIT for achieving efficient HOM interference in the frequency domain. Our study from the perspective of quantized light fields has provided a deeper understanding of the quantum properties of the double-$\Lambda$ medium. Our findings indicate that this scheme can perform high-fidelity Hadamard gate operations on frequency-encoded single-photon qubits and generate HOM two-photon NOON states with a fidelity greater than 0.99 at OD = 500. The proposed scheme offers high visibility and practical applications in photonic quantum metrology. Moreover, we have shown that the scheme can be used to realize arbitrary single-qubit gates and two-qubit SWAP gates by controlling the laser detuning and phase, which highlights its multifunctional properties. These photonic gates can be easily integrated with various EIT-related quantum technologies, such as quantum memories~\cite{QM1, QM2, QM3, QM4, QM5, QM6} and quantum repeaters~\cite{QR1, QR2, QR3, QR4, QR5, QR6, QR7}, enabling the development of quantum networks and scalable quantum information processing~\cite{QN1, QN2, QN3, QN4, QN5}. Overall, our results demonstrate that the proposed double-$\Lambda$ FWM system is a promising platform for frequency-domain HOM interference with high fidelity and multifunctional capabilities. Further experimental investigations and applications of this scheme in various quantum information processing tasks are warranted.


\section*{ACKNOWLEDGMENTS}

We thank Ite A. Yu, Ying-Cheng Chen, and Che-Ming Li for helpful discussions. This work was supported by the National Science and Technology Council of Taiwan under Grant No. 111-2112-M-006-027, No. 111-2639-M-007-001-ASP, and No. 111-2119-M-007-007.


\appendix

\section{COEFFICIENTS OF COUPLED EQUATIONS} \label{sec:Appendix A}

In Sec.~\ref{sec:Theory}, we derive the frequency-domain coupled equations between the probe field and the signal field, as shown in Eqs. (\ref{eq:(8)}) and (\ref{eq:(9)}). All coefficients in Eq. (\ref{eq:(10)}) can be obtained as follows:
\begin{align}
	\Lambda_p & = \Lambda_s = \frac{\alpha\Gamma}{4L}\left(\frac{1}{\Gamma + i\Delta}\right), \label{eq:(A1)} \\
	\kappa_p & = \kappa_s = -\frac{\alpha\Gamma}{4L}\left(\frac{1}{\Gamma + i\Delta}\right), \label{eq:(A2)} \\
	\zeta^p_{21} & = i\sqrt{\frac{\alpha\Gamma}{4L}}\left[\frac{i\Gamma - 2\Delta}{(\Gamma + i\Delta)|\Omega|}\right] e^{-i\phi_c}, \label{eq:(A3)} \\
	\zeta^p_{31} & = -i\sqrt{\frac{\alpha\Gamma}{4L}}\left(\frac{1}{\Gamma + i\Delta}\right), \label{eq:(A4)} \\
	\zeta^p_{41} & = i\sqrt{\frac{\alpha\Gamma}{4L}}\left(\frac{1}{\Gamma + i\Delta}\right) e^{-i(\phi_c-\phi_d)}, \label{eq:(A5)} \\
	\zeta^s_{21} & = i\sqrt{\frac{\alpha\Gamma}{4L}}\left[\frac{i\Gamma}{(\Gamma + i\Delta)|\Omega|}\right] e^{-i\phi_d}, \label{eq:(A6)} \\
	\zeta^s_{31} & = i\sqrt{\frac{\alpha\Gamma}{4L}}\left(\frac{1}{\Gamma + i\Delta}\right) e^{i(\phi_c-\phi_d)}, \label{eq:(A7)} \\
	\zeta^s_{41} & = -i\sqrt{\frac{\alpha\Gamma}{4L}}\left(\frac{1}{\Gamma + i\Delta}\right). \label{eq:(A8)}
\end{align}


\section{NOISE CORRELATIONS AND DIFFUSION COEFFICIENTS} \label{sec:Appendix B}

In the derivation of Eq. (\ref{eq:(17)}), we assume that the Langevin noise operator satisfies the delta correlation as follows:
\begin{align}
	\langle \tilde{f}_{jk}(z,\omega)\tilde{f}_{k'j'}(z',\omega')\rangle 
	=\frac{L}{2\pi c}D_{jk,k'j'} \delta(z-z')\delta(\omega-\omega'), \label{eq:(B1)}
\end{align}
where $D_{jk,k'j'}$ is the diffusion coefficient, which can be obtained using the Einstein relation~\cite{GarrisonQO} as follows:
\begin{align}
	D_{jk,k'j'} & = \frac{\partial}{\partial{t}}\langle \hat{\sigma}_{jk}\hat{\sigma}_{k'j'} \rangle 
	- \left\langle \left( \frac{\partial}{\partial{t}} \hat{\sigma}_{jk}-\hat{F}_{jk} \right) \hat{\sigma}_{k'j'} \right\rangle 
	\notag\\
	& - \left\langle \hat{\sigma}_{jk} \left(   \frac{\partial}{\partial {t}} \hat{\sigma}_{k'j'}-\hat{F}_{k'j'} \right) \right\rangle
	. \label{eq:(B2)}
\end{align}
For the case of $\gamma_{21}=0$ considered in the main text, we can obtain the diffusion coefficient of the double-$\Lambda$ atomic system through Eq. (\ref{eq:(B2)}). Its matrix form is
\begin{align}
	D_{jk,k'j'} 
	& =
	\begin{bmatrix}
		D_{2112} & D_{2113} & D_{2114} \\
		D_{3112} & D_{3113} & D_{3114} \\
		D_{4112} & D_{4113} & D_{4114} 
	\end{bmatrix}
    \notag\\
	& =
	\begin{bmatrix}
		\frac{\Gamma}{2}(\langle \hat{\sigma}_{33} \rangle + \langle \hat{\sigma}_{44} \rangle ) & 0 & 0 \\
		0 & 0 & 0 \\
		0 & 0 & 0 
	\end{bmatrix}
	. \label{eq:(B3)}
\end{align}
Under the condition of the weak probe and signal fields, $\langle \hat{\sigma}_{33}\rangle \approx \langle \hat{\sigma}_{44}\rangle \approx 0$, that is, all the normal-ordered noise correlations $\langle \tilde{f}_{jk}\tilde{f}_{k'j'}\rangle$ are close to zero. According to Eq. (\ref{eq:(10)}), the normal-ordered noise correlations contribute additional photons to the probe and signal fields, i.e. $\langle\tilde{a}^{\dagger}_{p,L}\tilde{a}_{p,L}\rangle$ and $\langle\tilde{a}^{\dagger}_{s,L}\tilde{a}_{s,L}\rangle$, which can be regarded as the spontaneous-emitted noise photons induced by the vacuum field reservoir. Additionally, we calculate the anti-normal-ordered noise terms, $\langle \tilde{f}_{k'j'}\tilde{f}_{jk}\rangle$. Under the same conditions, the corresponding diffusion coefficient matrix is
\begin{align}
	D_{k'j',jk} 
	& =
	\begin{bmatrix}
		D_{1221} & D_{1231} & D_{1241} \\
		D_{1321} & D_{1331} & D_{1341} \\
		D_{1421} & D_{1431} & D_{1441} 
	\end{bmatrix}
    \notag\\
    & \approx
	\begin{bmatrix}
		0 & 0 & 0 \\
		0 & \Gamma\langle \hat{\sigma}_{11} \rangle & 0 \\
		0 & 0 & \Gamma\langle \hat{\sigma}_{11} \rangle
	\end{bmatrix}
	. \label{eq:(B4)}
\end{align}
Unlike the normal-ordered noise terms, Eq. (\ref{eq:(B4)}) indicates that the anti-normal-ordered noise correlations in this double-$\Lambda$ system do not disappear even in the case of weak probe and signal fields. In Appendix~\ref{sec:Appendix C}, we will show that the anti-normal-ordered noise terms lead to the dissipation of the incident light fields.


\section{COMMUTATION RELATIONS OF FIELD OPERATORS} \label{sec:Appendix C}

The frequency-domain commutation relation of the creation and annihilation operators for the probe field can be expressed as
\begin{align}
	\left[\tilde{a}_p(z,\omega),\tilde{a}^{\dagger}_p(z,\omega')\right] = \frac{L}{2\pi c} \delta(\omega-\omega'). \label{eq:(C1)}
\end{align}
We use Eq. (10) in Sec.~\ref{sec:Theory} to calculate the communication relation of the probe field propagating through the double-$\Lambda$ atomic medium as follows:
\begin{align}
	\langle 
	\tilde{a}^{\dagger}_{p,L}\tilde{a}_{p,L}\rangle & = |A^{\ast}|^2 \langle \tilde{a}^{\dagger}_{p,0}\tilde{a}_{p,0}\rangle +|B^{\ast}|^2 \langle \tilde{a}^{\dagger}_{s,0}\tilde{a}_{s,0}\rangle 
	\notag\\
	& + \frac{L}{2\pi c}\sum_{jk}\sum_{j'k'}\int^{L}_{0}\mathcal{P}_{jk}D_{jk,k'j'}\mathcal{P}^{\ast}_{j'k'}dz, \label{eq:(C2)} \\
	\langle \tilde{a}_{p,L}\tilde{a}^{\dagger}_{p,L}\rangle & = |A^{\ast}|^2 \langle \tilde{a}_{p,0}\tilde{a}^{\dagger}_{p,0}\rangle +|B^{\ast}|^2 \langle \tilde{a}_{s,0}\tilde{a}^{\dagger}_{s,0}\rangle 
	\notag\\
	& + \frac{L}{2\pi c}\sum_{jk}\sum_{j'k'}\int^{L}_{0}\mathcal{P}^{\ast}_{j'k'}D_{k'j',jk}\mathcal{P}_{jk}dz. \label{eq:(C3)}
\end{align}
Substituting Eqs. (\ref{eq:(C2)}) and (\ref{eq:(C3)}) into Eq. (\ref{eq:(C1)}), and combining the results of Eqs. (\ref{eq:(B3)}) and (\ref{eq:(B4)}), the expectation value of the commutator of the output probe field is derived as
\begin{align}
	\left\langle[\tilde{a}_{p,L},\tilde{a}^{\dagger}_{p,L}]\right\rangle 
	& = \frac{L}{2\pi c} \left(|A^{\ast}|^2 +|B^{\ast}|^2\right)
	\notag\\ 
	+ & \frac{L}{2\pi c} \int^{L}_{0}(\mathcal{P}^{\ast}_{31}D_{1331}\mathcal{P}_{31}+\mathcal{P}^{\ast}_{41}D_{1441}\mathcal{P}_{41})dz. \label{eq:(C4)}
\end{align}
Based on Eqs. (C1) and (C4), we obtain
\begin{align}
	\int^{L}_{0}(\mathcal{P}^{\ast}_{31}D_{1331}\mathcal{P}_{31} & + \mathcal{P}^{\ast}_{41}D_{1441}\mathcal{P}_{41})dz
	\notag\\ 
	& =  1-|A^{\ast}|^2 -|B^{\ast}|^2. \label{eq:(C5)}
\end{align}
The left-hand side of Eq. (\ref{eq:(C5)}) comes from the contribution of anti-normal-ordered noise correlations. This indicates that once the output probe field is dissipated, i.e. $|A^{\ast}|^2 +|B^{\ast}|^2 < 1$, it must be caused by the anti-normal-ordered noise correlations. However, for a double-$\Lambda$ medium with a sufficiently large OD, $|A^{\ast}|^2 +|B^{\ast}|^2 \approx 1$, that is, the contribution of anti-normal-ordered noise terms approaches to zero, indicating that the vacuum field reservoir is unable to disturb the lossless system through quantum fluctuations. Additionally, the formula for the output signal field has also been derived as follows:
\begin{align}
	\int^{L}_{0}(\mathcal{Q}^{\ast}_{31}D_{1331}\mathcal{Q}_{31} & + \mathcal{Q}^{\ast}_{41}D_{1441}\mathcal{Q}_{41})dz
	\notag\\ 
	& =  1-|C^{\ast}|^2 -|D^{\ast}|^2. \label{eq:(C6)}
\end{align}
Note that in the above derivations, the  coefficients $\mathcal{P}_{jk}$ and $\mathcal{Q}_{jk}$ are in the form of
\begin{align}
	\begin{bmatrix}
		\mathcal{P}_{jk} \\
		\mathcal{Q}_{jk}
	\end{bmatrix}
    =e^{-M(L-z)}
    \begin{bmatrix}
    	\zeta^p_{jk} \\
    	\zeta^s_{jk} \\
    \end{bmatrix}
    ,
    \label{eq:(C7)}
\end{align}
where $\zeta^p_{jk}$ and $\zeta^s_{jk}$ can be obtained from Eqs. (\ref{eq:(A3)})-(\ref{eq:(A8)}).



\end{document}